\documentstyle[12pt]{article}
\def\beq{\begin{equation}}   \def\eeq{\end{equation}}
\begin{document}
\begin{titlepage}

\begin{flushright}
TPI-MINN-00/03-T\\
UMN-TH-1836/00\\
NYU-TH-00/01/01\\
hep-th/0001072\\
\end{flushright}

\vspace{0.3cm}

\begin{center}
\baselineskip25pt

{\Large\bf Families as Neighbors in Extra Dimension}

\end{center}

\vspace{0.3cm}

\begin{center}
\baselineskip12pt

{\large G. Dvali}

\vspace{0.2cm}
Department of Physics, New York University, New York, NY 10003

\vspace{0.2cm}

{\em and}

\vspace{0.3cm}
{\large  M.~Shifman} 

\vspace{0.2cm}
Theoretical Physics Institute, University of Minnesota, Minneapolis, 
MN 55455

\vspace{1.5cm}

{\large\bf Abstract} 

\vspace*{.25cm}

\end{center}

We  propose a new mechanism for explanation of the fermion 
hierarchy without introducing any family
symmetries. Instead, we postulate that different generations live on 
different branes embedded in a relatively large extra dimension, 
where gauge fields can propagate.  The electroweak symmetry is 
broken on a separate brane,
which is a source of exponentially decaying Higgs profile in the bulk.
The resulting fermion masses
and mixings are determined by an exponentially suppressed overlap 
of the fermion and Higgs wave functions and are automatically 
hierarchical even if  all copies are identical and there is no hierarchy 
of distances. In this framework the well known pattern  of  
the ``nearest 
neighbor mixing" is predicted due to the fact that the
families are literally  neighbors in the extra space. This picture may 
also
provide a new way of a hierarchically weak supersymmetry 
breaking,
provided that the combination of three family branes is a non-BPS 
configuration,
although each of them, individually taken, is.
This results in exponentially weak supersymmetry breaking. We also 
discuss
the issue of embedding identical branes in the compact spaces and
localization of the fermionic zero modes.

\end{titlepage}

\section{Introduction}

Large extra dimensions may help in understanding the hierarchy 
between the 
Planck and weak scales~\cite{add}. In the present paper we will 
concentrate
on the hierarchy of the fermion masses which is another mystery in 
the 
standard model (SM).  One possible approach
relies on (spontaneously broken) flavor symmetries.  But this does 
not really answer the question,
 rather  brings it at a different level.  
Instead of  explaining the hierarchy of the  Yukawa couplings, now 
one has to
explain the hierarchy of the breaking scales. 

 In the present paper we will
adopt a different attitude.  We assume that three SM
families are identical,   the difference in their masses is simply
because they happen to  live in different places in the extra space.  
More precisely, we assume that the original higher dimensional 
theory  admits, as its solution,  a brane with localized fermions with 
quantum numbers of one SM generation.
Multiple brane states will then generate $\nu$ identical copies of 
fermions, $\nu$ generations of the standard model.
Due to obvious reasons we will take $\nu =3$ in our discussion. It is 
clear that if the quarks and leptons
are to come from different branes,  then the gauge fields must freely
propagate in the interbrane space.
The transverse volume covered by gauge fields may be a 
world-volume of a ``fatter" brane,
or simply a compactified dimension. In any case there is an upper 
bound on a linear scale associated with this
 volume, $L\sim 1/(1$TeV). 

The crucial question in this picture is where does the electroweak
symmetry breaking happens?  We postulate that the vacuum
expectation value (VEV)  of the Higgs field
is induced due to the presence of a separate brane. The latter
acts as a source for the Higgs VEV
in the perpendicular direction, so that the Higgs VEV
 decays exponentially away
from the source,
\begin{equation}
H\sim {\rm e}^{-r/r_0}\,,
\end{equation}
where $r$ is  the distance from the source brane.

Thus, there is a nonzero Higgs profile in the bulk, and this will 
generate
masses of the SM fermions localized on  other branes.  In this
way the mass of the SM fermions will be determined by 
the overlap of its wave function (squared) with the Higgs profile. 
This  can be of order one for the
nearest brane, but exponentially suppressed for more distant 
neighbors.

Note that there is no need to postulate a hierarchy of distances
between immediate neighbors.  In any case, the hierarchy of the 
fermion masses is guaranteed.
Note also that there is an inevitable correlation between 
the masses and mixings.
The mixing between the fermions is suppressed by the
overlap of two (distinct) wave functions
with the Higgs profile. As a result the nearest neighbors
will mix  stronger than the next-to-nearest, and so on.
This pattern is well-known experimentally.

Other input assumptions are more or less standard
for the approach with the large (compact) extra dimensions.
Yet it is worth discussing them in brief.

We will consider one extra dimension,
so that the space has the topology of $M_4\times S$.
The size of the extra dimension $L$ is assumed to be much larger
than $M_{\rm Pl}^{-1}$ and the brane width $\delta$. Gravity is 
weak at these distances,
and plays essentially no role provided there are other
(passive) extra dimensions in which our construction is 
embedded.~\footnote{The
graviphoton is eliminated by the overall zero mode associated with
the breaking of the translational invariance in the fifth
direction.} A microscopic Planckean theory descends to distances
$L$ in the form of some field theory and the given geometry of 
space-time. This field theory is responsible for the
build-up of the branes, with the zero modes as discussed above.
The field(s) that ``build" the walls
are distinct from the matter and Higgs fields,
they have to be introduced for the wall-building purpose.

  Presumably, the characteristic size of our extra dimension,
should be somewhat smaller than inverse TeV, due to 
reasons associated with
the flavor
violation. In the theories with a low fundamental scale 
($M_{\rm Pf}$), 
there is a potential danger of higher-dimensional operators that
lead to a flavor violation in the low-energy processes
through the higher dimensional operators suppressed by powers of 
$M_{\rm Pf}$.
These can  be, in principle, controlled by gauging {\it non-Abelian}
flavor symmetries in the bulk~\cite{bd}. Flavor-violating exchange 
by  the bulk
flavor gauge fields or by the scalar flavons can be 
adequately suppressed \cite{bd}.
In our framework, however, there can be an
additional source of 
flavor violation
due to the exchange of the ordinary gauge fields~\cite{alex}.
\footnote{This is in contrast to other (unbroken) global symmetries
of 
the standard model (such as the baryon number), which, in principle,
 can be
protected by separating quarks from leptons in the extra space~\cite{nm}.}
This exchange is suppressed by the size of extra dimension versus
the localization width of the fermions, and may require the
size of extra 
dimension
to be below $1/(1000$TeV). In our discussion,
we will  assume
this bound to be satisfied, and keep the size as a free parameter.
Of course, making $L$ small implies increasing the cutoff of the 
theory,
and at some point it  will reintroduce the hierarchy problem. 
Therefore,  the issue of the low-energy supersymmetry (broken at
the  scale $\ll M_{\rm Pf}$) 
may become important in  our framework .

 In this respect it is interesting that localization of families on
the identical branes may automatically lead to a novel mechanism of
the {\it exponentially weak} supersymmetry breaking, provided 
that each individual
 family brane is a BPS state, whereas their combination is 
not. (Supersymmetry breaking on non-BPS branes was suggested in
\cite{ds1}.)
In Sec.~6 we will formulate a general sufficient condition for such a 
breaking.

Before proceeding  to a more detailed consideration 
let us summarize  crucial differences of  our
scenario from the existing alternative
high-dimensional mechanisms of the fermion mass generation.  In
\cite{nimasavas} the hierarchy of the
fermion masses was generated by invoking global flavor symmetries 
broken on a set of distant branes.
Each of the distant branes was responsible for the breaking of a 
particular subgroup of the full flavor group. This breaking then was 
communicated (``shined'') to the standard model fermions via a set of  
bulk 
messenger  fields, in some representation of the flavor group. 
Although the SM fields were
localized on the same brane, the resulting pattern of flavor 
symmetry breaking can still be hierarchical if the branes responsible 
for different breakings are located at  different distances.

Note that in this picture it is essential to have a flavor symmetry, as 
well as a set of the flavor-breaking branes with a variety of sets of  
the VEV's, plus a sector of the bulk messenger fields charged under 
the flavor group.  In our scenario there is no need to postulate any 
flavor symmetry at all. No messenger fields are needed. The only 
Higgs field that  acquires an 
expectation value is the standard model electroweak Higgs.  It is
impossible to avoid the hierarchical pattern of fermion masses, 
except for the unlikely case when all the three branes are stabilized 
right on top of each other. As will be discussed below, such 
stabilization is very difficult to achieve in practice, unless an 
unnatural distinction among the family branes is introduced.

 \section{Fermion Hierarchies from extra dimensions}

In this section we will discuss some model-independent features of 
the fermion masses in our framework and show  why a hierarchical 
pattern is inevitable.  Although this will not be crucial for our 
purposes, for simplicity and economy, we assume that all the 
standard model fermions are
generated from a single progenitor family in the original 
five-dimensional
theory. The corresponding  Yukawa couplings in the five-dimensional 
action are
\begin{equation}
 S = \int d^{5}x g_fH\bar f f_c
\end{equation}
where $H$ is the five-dimensional Higgs field and  $f$ and $f_c$ are
the five-dimensional fermions which give rise to the  
four-dimensional 
chiral $SU(2)\otimes 
U(1)$-doublet and singlet fermions, respectively.

Now, the fact that the SM fermions are localized zero modes on the 
branes, means that the five-dimensional fermionic fields allow for 
the expansion  of the form
\begin{equation}
f = \sum_i\Omega_i(y-y_i) f_i(x_{\mu}) + ...
\label{exp}
\end{equation}
where $y$ is the fifth coordinate and
$\Omega_i(y-y_i)$ are localized functions at  $y_i$, with
an exponentially  decaying profile. (It is assumed that
$|y_i-y_j|\ll L$.  At $|y-y_i|\sim (L/2)$ the exponential
decay regime changes, see below.)

The functions $f_i$ are zero modes of the four-dimensional Dirac 
operator. In this way the expansion (\ref{exp}) describes  
the zero-mode fermions of 
three standard model generations localized at the hyperspaces 
$y=y_i$ in the bulk. 

However, because of the finite distance 
between the branes these states are not 
completely orthogonal -- there is a nonzero overlap between
the  wave 
functions. This amounts to  a
nonzero but small mixing between the fermions.  What is interesting, 
there is a nontrivial correlation between the fermion mixing and 
their  masses.
This is clear from the way they are generated. The source of the 
masses is
the expectation value of the Higgs field, which we assume is induced 
on a separated ``source" brane, located
at some point $y=0$.  The Higgs VEV is maximal at the brane and 
decays exponentially in the bulk. Outside the 
Higgs VEV-generating brane, i.e. at $|y|>\delta$
\begin{equation}
H(y) = v \left(2e^{L/2}\right) {\rm cosh} \left[a\left( y - \frac{L}{2} 
\right)\right]
\label{profile}
\end{equation}
where $a$ is the mass scale defining the  inverse brane ``thickness"
and $v$ is the Higgs VEV on the brane.
Assuming that all three 
family branes
sit in the domain of the exponential decay of $H(y)$, 
the
fermion masses will be determined
by the overlap of the fermion wave functions with 
the exponential Higgs tail in 
the extra dimension, and, thus,
by the distance from the source brane.  Consider 
a situation when the  family
branes are located on the
same side from the source brane
(i.e. the Higgs VEV-generating brane). 
Then, even if they are placed in 
equal intervals, the hierarchical pattern
of masses  and mixings is guaranteed.  

For illustrative purposes we 
can approximate fermionic wave-functions by exponential profiles
\begin{equation}
\Omega_i= {\rm exp}\left( -|y - y_i|b\right)\,,
\end{equation}
where as in the case of the Higgs field, $b$ is a mass scale that sets
the ``thickness" of the fermionic profile(s), and we
 assume that all fermionic branes are identical (we neglect 
a small distortion of the wave functions
because of the non-zero overlap). Then, the fermion masses are given 
by the following overlap integrals:
\begin{equation}
m_{ij} = \int dy \, v{\rm exp}
\left\{-b\left( |y|{a\over b} + |y - y_i| + |y -
y_j|\right)\right\}\,.
\end{equation}

Note that,  equivalently,  we could have obtained the same result by 
going to the effective low-energy picture. The existence of the Higgs 
profile (\ref{profile}) means that there is a  four-dimensional Higgs 
state localized on the brane. This mode corresponds to  vibrations of 
the condensate and, therefore, is localized on the source brane.  
Thus, integrating out the extra dimension we will be left with a 
SM-like pattern, with the hierarchically suppressed Yukawa 
interactions.

\section{Higgs profiles}

In this and the next sections 
we will consider  some technical details such as the
generation of the Higgs condensate on a brane as well as the 
embedding of the
multiple fermionic branes in the compact dimension.
The generation of the Higgs condensates on the brane is a frequent 
phenomenon,
whenever there is a nontrivial coupling of the bulk scalars with the
brane.  For instance,  we will consider a simple example of the 
domain wall studied in \cite{ds}.  In this example there is a domain 
wall created  by a
real scalar field $\chi$,   and another field $H$ charged under the 
gauge group $G$.  In our case this will be assumed to be electroweak
 $SU(2)\otimes U(1)$.  It is essential that  the Lagrangian contains
interaction among these fields. In the simplest form the (scalar 
sector) of the five-dimensional  action can be taken as
\begin{equation}
 S = \int d^5x \,
\left[ |D_{\mu}H|^2 + |\partial_{\mu}\chi|^2 -  \left ( a(\chi^2 -
\mu^2)^2 +  (b\chi^2 -m^2)|H|^2 + |H|^4 \right)\right]\,.
\end{equation}
For $b\mu^2 > m^2$ the system has two ground states, with $\chi = 
\pm
\mu$ (and $H=0$ for both).
Due to simple topological arguments there is a wall (brane) 
interpolating
between the two.  For $H=0$
the wall profile has a simple form
\begin{equation}
\chi = \mu {\rm tanh}(y\mu\sqrt{4a})\,.
\end{equation}
Since $\chi$ goes through zero in the middle of the
wall, $H$ can become unstable and 
condense
on the brane.
This can be simply seen by examining  small perturbations  $H = 
H{\rm
e}^{i\omega t}$ in the wall background. The linearized Schr\"odinger
equation takes the form
\begin{equation}
\partial^2_y H - \left(
b\mu^2{\rm tanh}^2(y\mu \sqrt{4a}) - m^2\right)
H = \omega^2 H
\end{equation}
which clearly has an unstable eigenmode (with imaginary $\omega$) 
for some range of parameters.

Thus, the SM symmetry is spontaneously broken on the brane but 
is restored in
the bulk in the infinite volume limit.  For the finite extra dimension,
the gauge fields will get nonzero masses by interacting with 
the brane  condensate,
and the gauge symmetry will be spontaneously broken in 
the effective low-energy theory. 

Finally, let us remark on a technical point regarding the embedding 
of a  brane (or several branes, as opposed to the antibrane)
in the compact extra dimension. 
This issue was discussed in great detail in
\cite{HouLosevMS}.
The wall one may have on the cylinder is slightly different from the 
standard kink in the non-compact space, which 
interpolates between distinct vacua of the theory.
To have walls on  the cylinder
one must assume that the fields of which the walls are built
are defined on manifolds with noncontractible cycles.
For instance assume that $\chi$ is an angular variable
(an ``axionic" type field), defined modulo $2\pi$. The 
appropriate interaction 
potential is
\begin{equation}
V = 2M^4{\rm cos}\chi + 
(b {\rm sin}\chi - m^2)|H|^2 + |H|^4
\end{equation}
The resulting
brane is a sine-Gordon soliton
\begin{equation}
\chi = 4{\rm  tan}^{-1}{\rm exp}(ym) \,,
\label{solit}
\end{equation}
and can be embedded in the compact space. Since $\chi$ changes by 
$2\pi$
through the soliton, sin$\chi$ becomes zero and destabilizes $H$,
much in the same way as for the kink.

Moreover,  as it was shown in \cite{HouLosevMS},
the isolated wall on the cylinder may be BPS saturated,
i.e. leading to supersymmetric low-energy theory
of the zero modes. 
But we have several branes: three ``fermion" and 
one Higgs VEV-generating. 
It is natural to assume that, being considered in
isolation, each wall is BPS saturated. Taken all together
they
need not necessarily be BPS. Hence, one can get 
a supersymmetry breaking exponentially small in the parameter
$|y_i-y_j|/\delta$ where
$\delta$ is the wall ``thickness." In this way,
one gets an exponential suppression of the SUSY breaking scale 
without any
input hierarchy. 

To reiterate,  had we just one generation, 
supersymmetry will be unbroken.
It is the intergenerational (interbrane) interference
that makes the  wall configuartion
non-BPS, and breaks SUSY. 

This effect is  independent of the
other arguments presented above regarding
the fermion mass hierarchy.  
The original masslessness of the fermions is not  due to SUSY,
but, rather, due to the topological property of the brane (or
due to  a mechanism to be  discussed in Sec. 4). 
Note that the matter fields need not be chiral
at  distances $\ll L$, when our intermediate field theory flows
to a fundamental one.
The chirality of the trapped zero modes occurs
as a result of the winding of the
solution under consideration in the extra dimension.
In this picture it is
natural
that the matter fermions are lighter than the sfermions.

\section{Multiple branes in compact spaces}

  The fermionic fields must be localized on a
number of identical stable
branes, admitting the fermionic zero modes.  Stability of such  
branes, in
general, is due to some charge $Q$. This may be either a
topological  charge (e.g. in the case of the kink or soliton) or a charge 
with
respect to some higher forms
(e.g. as in the case of $D$ branes). The corresponding flux then
guarantees the stability of the
brane.  The same flux conservation then often forbids the embedding 
of the
branes in the compact space, since the flux can end nowhere.  One is 
then
forced to introduce antibranes on which the flux lines may end,
to balance the total charge.  In our scenario we would
like to avoid antibranes.
 
One possible way out then is to consider
topological charges compatible with the compact boundary 
conditions, as 
above (see \cite{HouLosevMS}).
For instance, we can put arbitrary number of solitonic branes of the 
form
(\ref{solit}) in the compact space.

Here we will consider an alternative way for
dealing with this issue in the cases when the flux conservation is 
incompatible
with periodicity of the space. Let such  charge be $Q$.
Then, instead of introducing an antibrane with the charge $-Q$,
we can assume that the charge in
question is Higgsed. Then the 
flux lines will be absorbed by the ``medium," much in the same way 
as the  conductor
absorbs  the  electric flux.
 
Let us consider the issue of the stability of such a system.
As a prototype toy model consider a
brane which is a source of a massless scalar field $\phi$. The
corresponding coupling of $\phi$ to the world-volume of the brane is
\begin{equation}
Q\int \,
dx_{\alpha}\wedge 
dx_{\beta}\wedge dx_{\gamma}\wedge dx_{\delta} \,
\epsilon_{\alpha
\beta\gamma\delta}\, \phi\,.
\end{equation}
In this way the brane ``shines'' a massless scalar field. If there are no
lighter states in the theory charged under $Q$, the brane will be 
stable
due to the flux conservation. The massless field $\phi$ satisfies 
the classical
 equation with the delta-function source in the fifth coordinate $y$ 
(transverse to
the brane)
\begin{equation}
\partial^2_y \phi = Q\delta (y)\,. 
\end{equation}
The  solution of the above equation is evident,
\begin{equation}
\phi = Q|y|\,.
\end{equation}
This solution shows that it is impossible to put such field
on the cylinder. 

Imagine now that we
give a small mass $m$ to the scalar field. This will guarantee
that the
flux is screened at large distances $y \gg  m^{-1}$ as
\begin{equation}
\phi \sim Q{\rm exp}(-m|y|)\,.
\end{equation}
The exponential solution above is valid for the noncompact
fifth dimension. If it is compactified (a circle),
the solution with the appropriate boundary conditions rather takes 
the
form
\begin{equation}
\phi \sim Q{\rm cosh }\left[ m\left( y - \frac{L}{2}\right)\right]\,
\quad 0 \leq y \leq L \,.
\end{equation}
The existence of such a brane is  compatible with 
compactification.

Let us consider the issue of the fermionic zero modes on such 
branes.
Consider the following coupling of a bulk fermion to $\phi$
\begin{equation}
g \, (\partial_A\phi )\, \bar{\psi}\gamma_A\psi\,.
\end{equation}
This creates a $\theta(y)$ function type mass term, which chages 
the sign across
the brane
\begin{equation}
gQ\theta(y)\bar{\psi}\gamma_5\psi\,.
\end{equation}
Due to the index theorem there is a localized zero mode on the brane,
which at $|y| \ll L/2$ takes the form
\begin{equation}
\psi = f(x_{\mu}){\rm exp}(- g|y|Q)\,.
\end{equation}
This zero mode will
persist for the nonzero mass of the $\phi$ as well.

 Note that in the  above discussion we could have 
used antisymmetric $4$-form field
$A_{\alpha...\beta}$ instead of the scalar $\phi$. (Such fields are
present in many brane constructions
(e.g. see \cite{forms}). To the best of our knowledge, however, this is
the first attempt of using them for localizing the chiral fermionic
zero modes).
The whole discussion would go
through, except that the fermion couplings now would be modified as
follows. The coupling to the brane is
\begin{equation}
Q\int dx_{\alpha}\wedge dx_{\beta}\wedge dx_{\gamma}\wedge 
dx_{\delta}
\, A_{\alpha
\beta\gamma\delta}\,,
\end{equation}
and the fermions now  couple to the $5$-form field strength
\begin{equation}
 F_{\alpha\beta\gamma\delta\omega} = \partial_{[\alpha}A_{
\beta\gamma\delta\omega]}
\end{equation}
which changes the sign across the brane $F \sim \theta(y)$. The 
fermions
coupled to $F$
\begin{equation}
 F_{\alpha\beta\gamma\delta\omega}
\epsilon^{\alpha\beta\gamma\delta\omega}
\bar{\psi}\psi
\end{equation}
will develop a zero mode on the brane.

\section{Prototype Model}

 Now we are in a position to write down a simple prototype 
five-dimensional
Lagrangian giving
rise to the desired structure. The important interaction terms are
\begin{equation}
L = |\Lambda - X^3|^2  +  2M^4{\rm cos}\chi +  
(b {\rm sin}\chi - m^2)|H|^2 + |H|^4 \\
 +  X^2r_f\bar{f}f + X^{*2}r_{f_c}\bar{f_c}f_c
 +  g_fH\bar {f}f_c
\end{equation}
plus the standard kinetic and gauge terms.
Here $X$ is a gauge-singlet scalar that breaks $Z_3$ symmetry and 
produces three
walls on a compactified dimension. 
 $X$ changes the phase by $2\pi/3$ 
through
each of the walls and creates a single zero mode from each species of 
fermions
$f, f_c$. These fermions have quantum numbers of one SM 
generation.
Moreover,
$f$ stands for $SU(2)\otimes U(1)$-doublet, while $f_c$ stands for 
$SU(2)\otimes U(1)$-singlet
states, respectively. 
The Yukawa couplings
with $H$ and $X$ are assumed to have 
the $SU(2)\otimes U(1)$ doublet and 
singlet
structures, respectively.
 In conventional notations the zero modes coming from $f$ are 
the left-handed quark and lepton
doublets $Q,L$ and the ones coming from $f_c$ are 
the left-handed  {\em anti}singlets
$u_c,d_c,e_c$.
 The precise form of the interaction terms between $X$ and $\chi$ is 
not 
very important,
provided that the $\chi$-wall gets stabilized on top of one of 
the $X$-walls.
This can be achieved, for instance, by adding~\cite{manyf}
\begin{equation}
(b' {\rm sin}\chi)|X|^2
\end{equation}
with a positive $b'$.
The above Lagrangian reproduces all the desired features discussed
above. It has three identical branes compatible with periodic 
boundary 
condition, 
with one SM generation
localized per brane. Plus a separate brane that breaks 
the electroweak 
symmetry.

\section{Hierarchical SUSY Breaking from 
Multiple \\
Branes}

 In this section we will argue that 
the presence of  the identical ``family 
branes"
may result to an exponentially weak supersymmetry breaking, even 
though each
individual brane, in isolation, may be supersymmetry 
preserving
(i.e. BPS saturated). This is the case if the multiple-brane state is a 
non-BPS
configuration. The idea that the observable  SUSY breaking may 
be due
to the fact that we live on a non-BPS brane, was first 
put forward  in 
\cite{ds1}.
Here we show that in the present circumstances this breaking may 
be 
exponentially weak.

Before formulating a very general sufficient condition for such  weak 
supersymmetry
breaking, let us illustrate the main point in a toy example. We are 
looking for a model
that gives rise to a BPS brane, but in which 
the states with two (or more) such 
branes are
are non-BPS. The simplest 
model of this type is the one with the  spontaneously broken 
$R$ symmetry (the symmetry is $Z_N$). 
In four-dimensions the superpotential can be chosen as~\cite{dk}
\begin{equation}
  W = \Lambda X  - {cX^{N + 1}\over N + 1}
\end{equation}
where $X$ is a chiral superfield.

This theory has stable domain wall solutions across which the phase 
of 
$X$ changes by
$2\pi/N$. Clearly, having $N$ such domain walls is compatible with 
periodicity of the
transverse coordinate. Because the superpotential changes through 
the 
wall, this system
admits a nontrivial central extension~\cite{ds1}; it can be shown that 
the elementary
wall is BPS saturated. For $N \rightarrow \infty$, the corresponding 
solutions
can be found explicitly~\cite{dk}. However, the $N$-wall state 
on the cylinder is 
not BPS
saturated,
generally speaking.\footnote{Note, however, that the central charge 
can still be defined, it does not vanish
for $N$-wall states~\cite{j1}. Due to this reason, in particular,
the junction of $N$ domain walls (in the {\em non}compactified 
space) can be BPS saturated~\cite{j2, j1}.}
Thus, the $N$-wall state, breaks all  supersymmetries.

 Let us assume that the transverse coordinate is compactified on a 
circle 
of radius
$R\,\, (=L/2\pi )$. The equilibrium state in such a case corresponds 
to 
$N$ branes 
around the circle
at  equal distances between the neighbors. What is the strength of 
the 
resulting 
supersymmetry breaking?
 It is exponentially suppressed by the
inter-brane distance
\begin{equation}
    \sim e^{-Rm}
\end{equation}
where $m$ is a mass of the $X$ quanta, the scale that sets the width 
of 
the brane
(we ignore factors of order $N$, which may be important, however, 
in the
large $N$
case). This weakness is not difficult to understand. The wall is a field 
configuration,
that approaches the vacuum state {\it exponentially rapidly} in the 
transverse coordinate.
Thus, unless there are massless fields that can ``carry away'' the 
message 
about its
presence, all the influence of the brane is exponentially suppressed 
at 
large distances.
So is the resulting supersymmetry breaking. 

 This gives us a very general sufficient condition for exponentially 
suppressed
supersymmetry breaking:

1) the presence of the BPS brane, whose stability
is {\it not} due to massless fields in the theory;

2) the $N$-brane states should not be BPS saturated.

Note that it is very important that the stability of the brane is due to 
the topological charge,
which is {\it not} a source of any massless bulk field.
If this stability were due to some other charge coupled to some 
massless 
bulk field
(e.g. as in the $D$ brane case), the corresponding field would serve as 
a 
messenger between the branes, and the resulting SUSY breaking 
would 
be power  suppressed.

 This is the crucial difference 
which differentiates  our mechanism from the 
conventional 
schemes, in which the SUSY breaking gets transmitted between the 
branes 
by some bulk messenger interactions (e.g. see \cite{rs},\cite{hw}). 

Once again, it is important to understand that massless fields in 
question 
are
those coupled to the stabilizing charge, and not other massless fields.
For instance, in any realistic theory, there is at least one massless 
field, 
the graviton,
coupled to the energy-momentum tensor of the brane. This, however, 
will not serve as
a messenger for the SUSY breaking, since individually all branes are 
SUSY
preserving, and
only their exponentially suppressed interaction brakes 
supersymmetry.

\vspace{0.5cm}

{\bf Acknowledgments}

 \vspace{0.1cm} 

We would like to thank Savas Dimopoulos, Gregory Gabadadze, Alex Pomarol and
Massimo Porrati for useful discussions.
A part of this work was done at ITP, Santa Barbara, where we were 
participants of the program ``Supersymmetric Gauge Dynamics and 
String Theory." We are grateful to the ITP staff for hospitality.
This work was supported in part by DOE under Grant No.
DE-FG02-94ER40823, by National Science Foundation under Grant 
No. PHY94-07194, and by David and Lucile Packard Foundation 
Fellowship for Science and Engineering.

\end{document}